\documentclass[10pt,aps,prl,twocolumn,floatfix, tightenlines,showpacs,showkeys,preprintnumbers,footinbib]{revtex4-1}
\pdfoutput=1
\usepackage{natbib}\usepackage[colorlinks=true,citecolor=blue,linkcolor=blue,breaklinks=true]{hyperref}
\usepackage{amsmath,amssymb,mathtools}
\usepackage{epsfig}  
\usepackage{graphicx}   
\usepackage{slashed}             
\usepackage{url}
\usepackage{color}
\usepackage{multirow}
\usepackage{placeins}
\usepackage[dvipsnames]{xcolor}

\allowdisplaybreaks

\bibpunct{[}{]}{,}{n}{}{,}

\newcommand{\ba}{\begin{array}}
\newcommand{\ea}{\end{array}}
\newcommand{\bd}{\begin{displaymath}}
\newcommand{\ed}{\end{displaymath}}
\newcommand{\beq}{\begin{equation}}
\newcommand{\eeq}{\end{equation}}
\newcommand{\bea}{\begin{eqnarray}}
\newcommand{\eea}{\end{eqnarray}}

\newcommand{\ra}{\rightarrow}

\def\g{\gamma}

\def\m{\mu}
\def\n{\nu}

\def\q2 {q^2}

\def\bt{\begin{table}}
\def\et{\end{table}}

\catcode`@=11 
\def \gsim{\mathrel{\mathpalette\@versim>}}
\def \lsim{\mathrel{\mathpalette\@versim<}}
\def \@versim#1#2{\lower0.4ex\vbox{\baselineskip\z@skip\lineskip\z@skip
     \lineskiplimit\z@\ialign{$\m@th#1\hfil##\hfil$%
     \crcr#2\crcr\sim\crcr}}}
\catcode`@=12 

\begin{document}

\title{Are We Looking at Neutrino Absorption Spectra at IceCube?}

\author{Siddhartha Karmakar}
\email{ phd1401251010@iiti.ac.in}
\author{Sujata Pandey}
\email{phd1501151007@iiti.ac.in}
\author{Subhendu Rakshit}
\email{rakshit@iiti.ac.in}
\affiliation{\em Discipline of Physics, Indian Institute of Technology Indore,\\
 Khandwa Road, Simrol, Indore - 453\,552, India}
 
\pacs{13.15.+g, 95.35.+d} 

\begin{abstract}
The observed spectrum of ultrahigh energy neutrinos at IceCube might be indicative of absorption of such neutrinos in ultralight dark matter halos. We point out that various features of this spectrum can be explained by such absorptions. 
For a light $Z^\prime$-mediated $t$-channel interaction between dark matter and neutrinos, we propose a novel  mechanism of absorption of these neutrinos at particular energies. This can save the models of AGN predicting large neutrino flux at energies more than a PeV. 
\end{abstract}

\maketitle

Recently, IceCube collaboration succeeded~\cite{IceCube:2018dnn} in pointing back to a specific blazar TXS 0506+056 as the source of one ultrahigh energy (UHE) neutrino event observed at IceCube, located at the South Pole. With this begins a new chapter of multi-messenger astronomy. But many of the models, proposed to describe the dynamics of blazars, predict large neutrino flux after a PeV~\cite{Murase:2010gj} which is disfavored by the observed spectrum at IceCube: Very few events have been observed above a PeV with no event that could correspond to the Glashow resonance. This is suggestive of a sharp cut-off in the spectrum around a PeV. Between 400 TeV to 1 PeV, very few events have been observed, implying an apparent depletion in the spectrum. The puzzle is further accentuated by an excess of events at energies between 60$-$160 TeV.

Various astrophysical objects are expected to be the source of such neutrinos, and for each one of these categories different dynamics might be in action, yielding different spectra.  We propose that various features of the observed spectrum can be the outcome of absorption of these spectra in ultralight dark matter\,(DM) halo pervading the cosmos. This kind of sub-eV DM can exist in the form of a non-thermal Bose-Einstein condensate (BEC)~\cite{Hu:2000ke} and is thus a cold DM candidate that can saturate the total relic density of DM. Such ultralight DM can help address issues pertaining to the structure formation in the early Universe, namely the cusp-core problem and the missing satellite problem, that predicts many satellite galaxies which contradicts with the observations. Fermionic ultralight DM are considered to be hot and hence cannot account for more than one percent of the observed relic density. Here we restrict ourselves to the scalar BEC DM of mass below an eV, as it ensures that the critical temperature of the condensate is greater than the temperature of the Universe at any epoch~\cite{Das:2014agf}.

The UHE neutrinos can interact with the DM $\phi$ {\it via} the exchange of a $Z^\prime$ vector boson which can be as light as 10~MeV:
\bea
\mathcal{L} &\supset & ig (\phi^{*} \partial_{\mu} \phi-\phi \, \partial_{\mu} \phi^{*}) Z^{\prime \m} + i f \bar{\nu}_{\tau} \g_{\m} \n_{\tau} Z^{\prime \m} \, .
\eea 
Here only the third generation leptons interact with the $Z^\prime$. Although only the $\nu_\tau$s get depleted by such $\nu$--DM interactions, as the neutrino oscillates to a different flavor more promptly than it can interact with a DM particle, the reduction of the number of neutrinos of a particular energy occurs uniformly for all flavor of neutrinos. The same happens for anti-neutrinos as well.  A detailed model building endeavor, justifying the choice of parameters and computation of cross-section, will be reported elsewhere~\cite{spsksr2}. 

The elastic scattering $\nu \phi \ra \nu \phi$ degrades the energy of the neutrinos, leading to regeneration of neutrinos at lower energy bins. Such a regeneration is quantified by the $Z$-factor that alters~\cite{Naumov:1998sf} the interaction length $\lambda=1/(n\sigma)$ to $\Lambda(E,X)=\lambda(E)/(1-Z(E,X))$. Here $n$ and $\sigma$ stand for the number density of the DM and $\nu$--DM cross-section respectively. The expression for $Z(E,X)$ can be located in Ref.~\cite{Rakshit:2006yi} where $E$ stands for the neutrino energy and $X$ is related to the traversed path-length. As a result, as neutrinos whizz through the DM cloud, the flux suffers an exponential depletion $F[E,X]= F[E,0] \exp[-X/\Lambda(E,X)].$ 

If some of these neutrinos are generated at a very early epoch of the Universe, then depending on the redshift,  the aforementioned degradation of neutrino energy may get further enhanced as around that era the number density of DM would have been higher. This leads to shorter interaction lengths of neutrinos travelling through the DM soup~\cite{Blum:2014ewa}, leading to frequent interactions that repeatedly degrade the energy of the neutrinos. We do not need to go into such complications to illustrate the ideas floated in this letter. We work with UHE neutrino sources located within 200\;Mpc from the Earth to get a conservative estimate of this degradation and take the density of the dark matter as $1.2 \times 10^{-6}$\;GeV/cc uniform across the Universe.

\begin{figure}[h!]
 \begin{center}
 \includegraphics[width=2.7in,height=2.0in, angle=0]{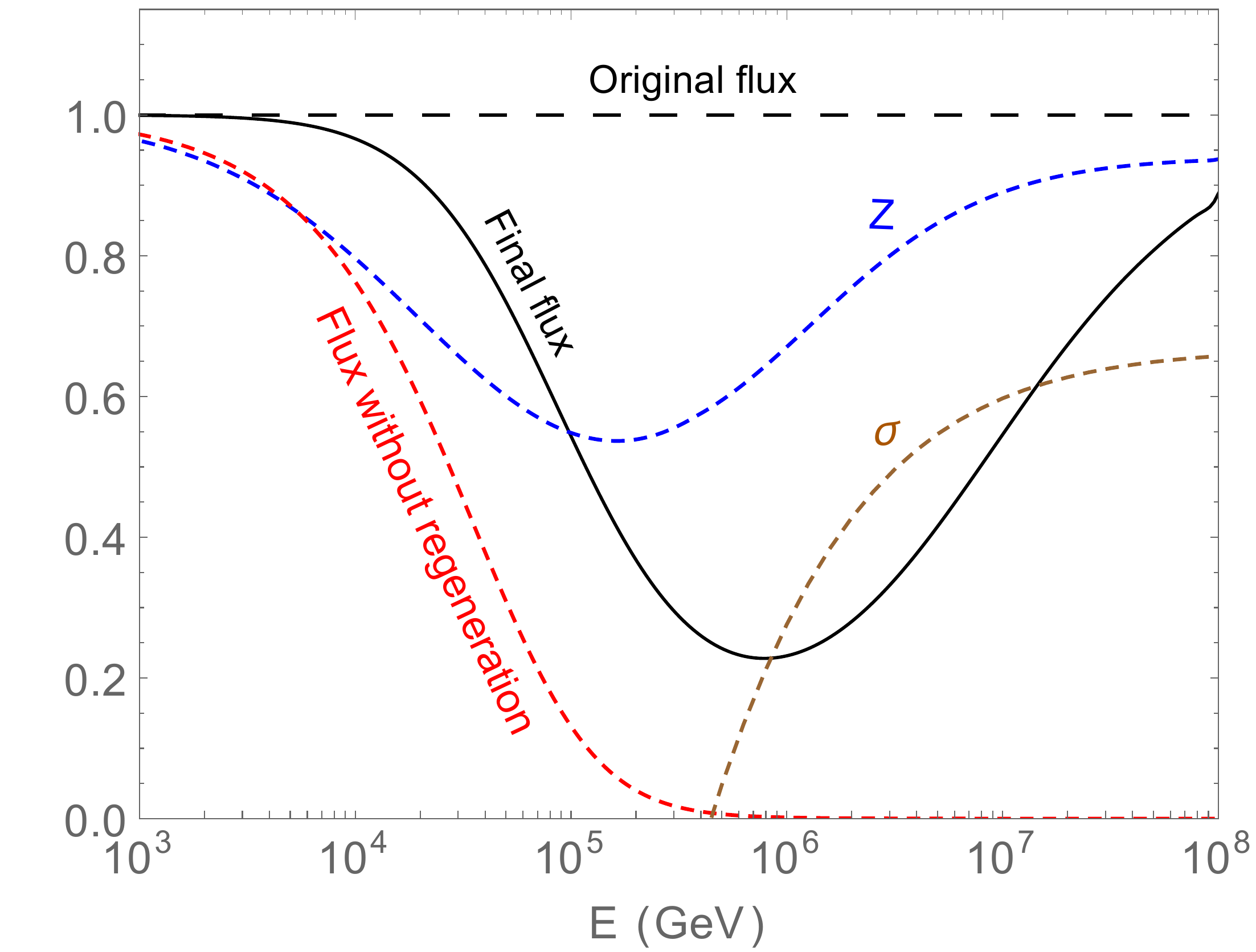}
 \caption{
The solid and dashed black lines represent the final and the original flux, multiplied by $E^2$, in units of $\rm{GeV~cm}^{-2}~\rm{s}^{-1}~\rm{str}^{-1}$ and are scaled by a factor $3 \times 10^{9}$. The same is true for dashed red line that  denotes the effect of attenuation alone. The blue-dashed line is the $Z$-factor responsible for regeneration. The brown-dashed line refers to the neutrino-DM scattering cross-section in units of eV$^{-2}$ and is scaled  by $3 \times 10^{21}$. Here $m_{\text{DM}} = 0.3$~eV and $g f = 7 \times 10^{-3}$. For all plots  $m_{Z^\prime} = 10$~MeV and $m_{\nu} = 0.1$~eV.}
 \label{fig:nocutoff}
 \end{center}
 \end{figure}
 We illustrate our proposed mechanism behind neutrino absorption in Fig.~\ref{fig:nocutoff} with an $E^{-2}$ neutrino flux. For the chosen benchmark parameters, the $t$-channel $\nu$--DM cross-section $\sigma$ rises sharply around a PeV and flattens after that. At lower energies, when $\sigma$ is not appreciable, both attenuation and regeneration are negligible, leading to $Z\sim 1$. At very high energies, when $\sigma$ flattens, the neutrinos lost due to attenuation get regenerated from the higher energy bins, leading again to $Z\sim 1$. In between these two extremes, $Z$ deviates from unity. As a result, after taking regeneration into account, the net attenuated flux shows a typical absorption spectrum, with a dip at intermediate energies. One may note that the effect of regeneration is quite pronounced at higher energies where the cross-section is sizeable. In all plots we have computed $Z$ up to the third order while ensuring its convergence. Consequently, the neutrino flux conservation is also guaranteed. 
 
\begin{figure}[h!]
 \begin{center}
 \includegraphics[width=2.7in,height=2.0in, angle=0]{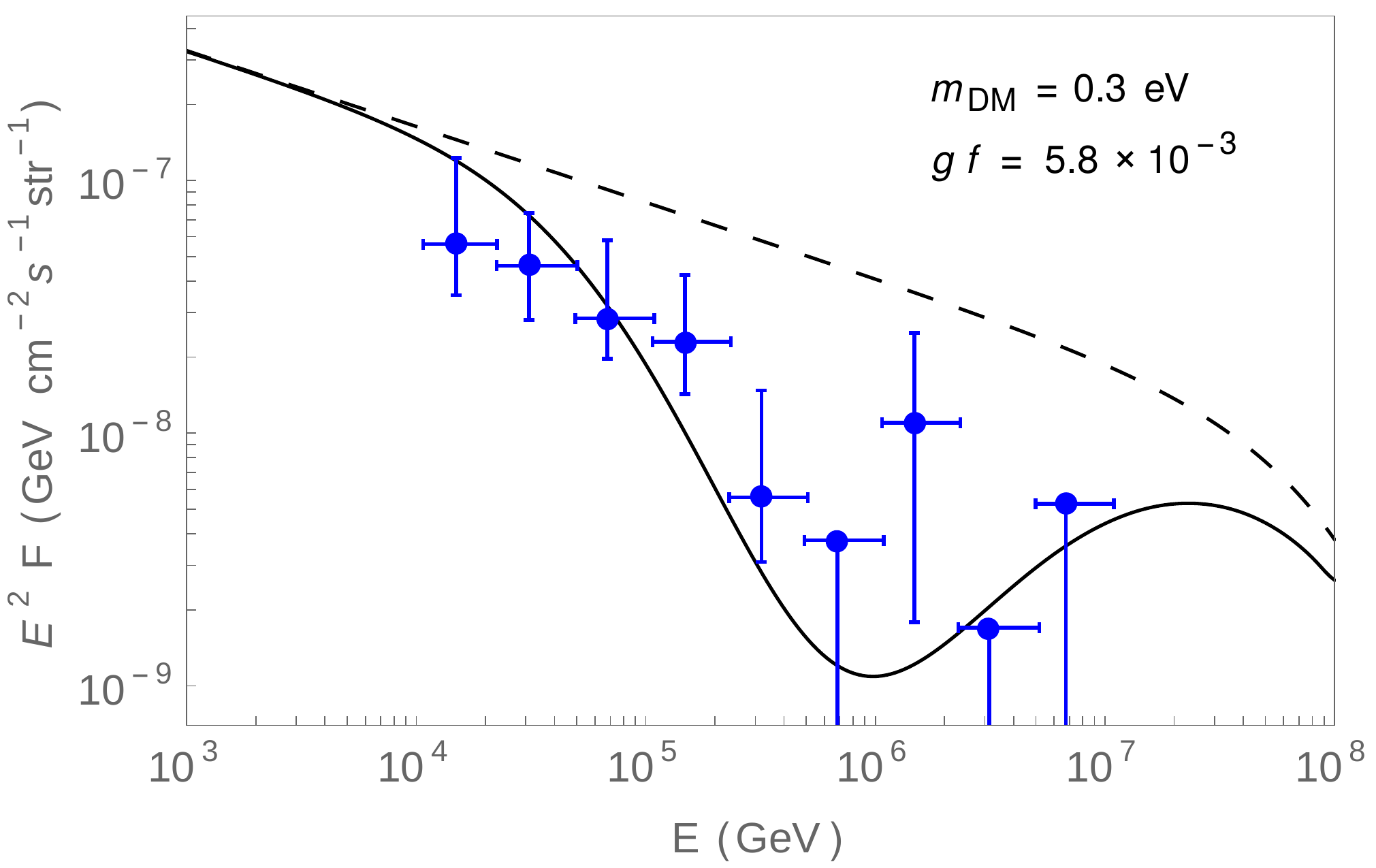}
 \caption{
Absorption dip due to $\nu$--DM interaction for a diffuse neutrino flux which follows the $E^{-2.3}$ power law. The solid and dashed black lines represent the attenuated flux  and the original flux, both multiplied by $E^2$. In the plots the blue points represent the fitted  diffuse astrophysical flux to the 7-year IceCube data~\cite{AlbrechtKarle}. }
 \label{fig:MBFlux}
 \end{center}
 \end{figure}

The above mechanism helps explain the shortfall of events above 400~TeV as shown in Fig.~\ref{fig:MBFlux} using an $E^{-2.3}$ flux with a normalisation of $8 \times 10^{-8}$ at $100$~TeV~\cite{Murase:2010gj} and an exponential cut-off at $E_{\nu}=100$ PeV. Note that we use a fixed flux only for the purpose of illustration. We do not intend to fit the data with our proposed method. Although the shortfall of events for neutrino energies higher than 400~TeV with only a few events around a PeV still persists, due to the limited statistics, it is still premature to comment on the shape of the dip.  

The novelty of our mechanism lies in the generation of an absorption dip {\it via} neutrino regeneration through a $t$-channel $\nu$--DM interaction. The shape of the dip results from an interplay of the energy dependences of the original neutrino flux and the cross-section. 
A sharper dip can be obtained with a steeper rise of cross-section with energy. Although we have used a $Z^\prime$ mediated interaction to illustrate this idea of neutrino absorption, it is possible to construct other models that can also give rise to appreciable $\nu$--DM interactions. The resulting energy dependence of the cross-section will be different leading to absorption dips of different shapes. In this $Z^\prime$ mediated model, the onset of the flattened part in the cross-section shifts to higher energies for lower masses of the DM. As the mass of the ultralight BEC DM cannot exceed $\sim 1$~eV, this onset cannot happen before 800~TeV or so, for mass of the mediator $m_{Z^\prime} = 10$~MeV. As below this energy the cross-section falls very fast, all the low energy phenomenology with neutrinos remain unaffected, including neutrino cosmology.

\begin{figure}[h!]
 \begin{center}
 \includegraphics[width=2.7in,height=2.0in, angle=0]{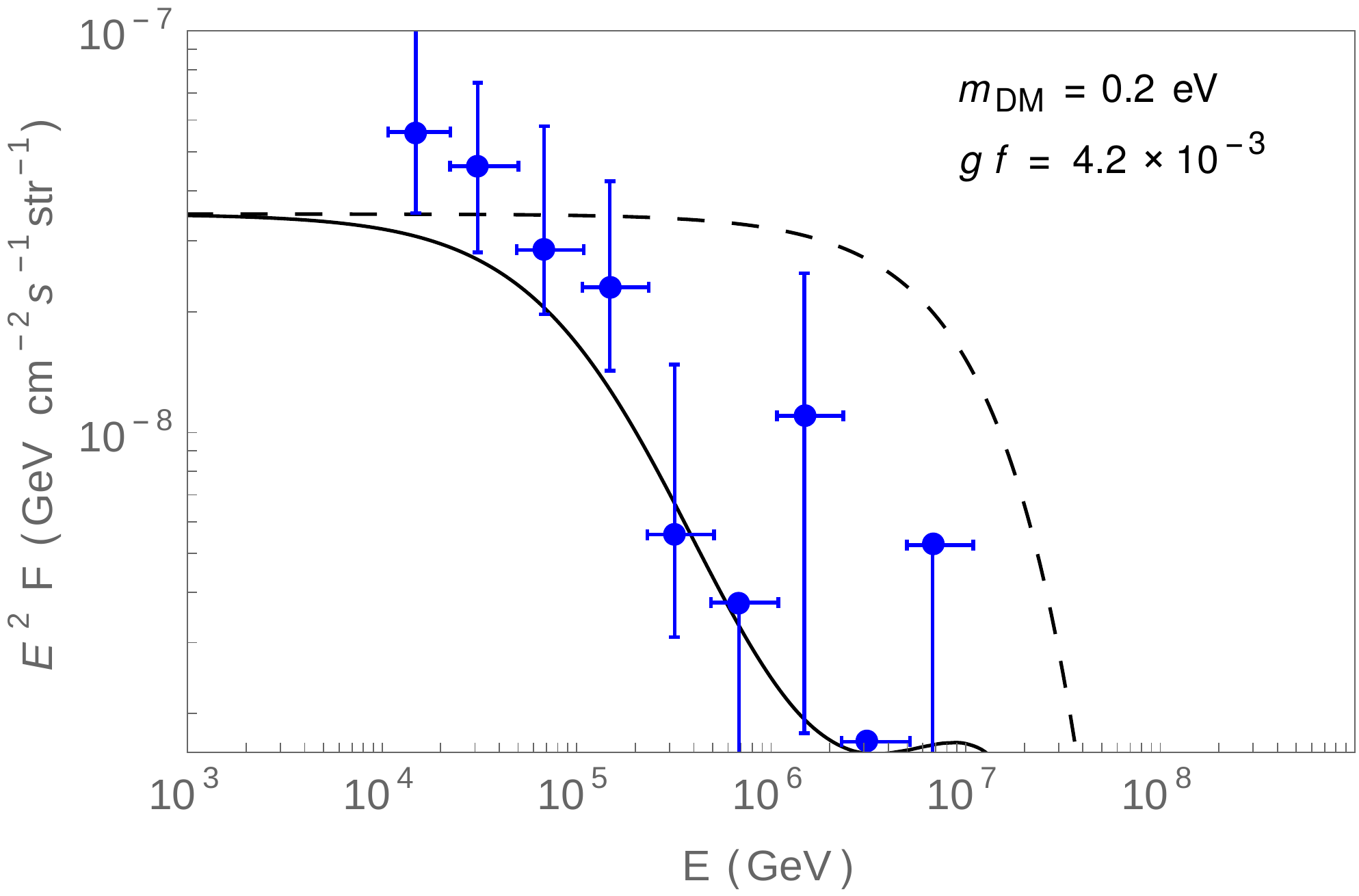}
 \caption{
An early cut-off in a diffuse neutrino flux due to $\nu$-DM interactions. The dashed line represents an $E^{-2}$ original flux with a cut-off at $12$ PeV. The solid line results from attenuation and regeneration of neutrinos due to such interactions. Both fluxes are multiplied by $E^2$ in the plot.}
 \label{fig:cutoff}
 \end{center}
 \end{figure}

With a suitable choice of model parameters the dip can be wider. This can be the origin for a sharp cut-off after a PeV as shown in Fig.~\ref{fig:cutoff} starting with a flux $F= 3.5 \times 10^{-8} ~ E^{-2} \exp[-E/(1.2\times 10^7 \,{\rm GeV})]$ GeV$^{-1}$ cm$^{-2}$ s$^{-1}$ str$^{-1}$. We note that for the chosen model parameters,  the cut-off recedes by almost two orders of magnitude, which is the outcome of repeated  interactions of the neutrinos with DM, on its way to the Earth.

\begin{figure}[h!]
 \begin{center}
 \includegraphics[width=2.7in,height=2.0in, angle=0]{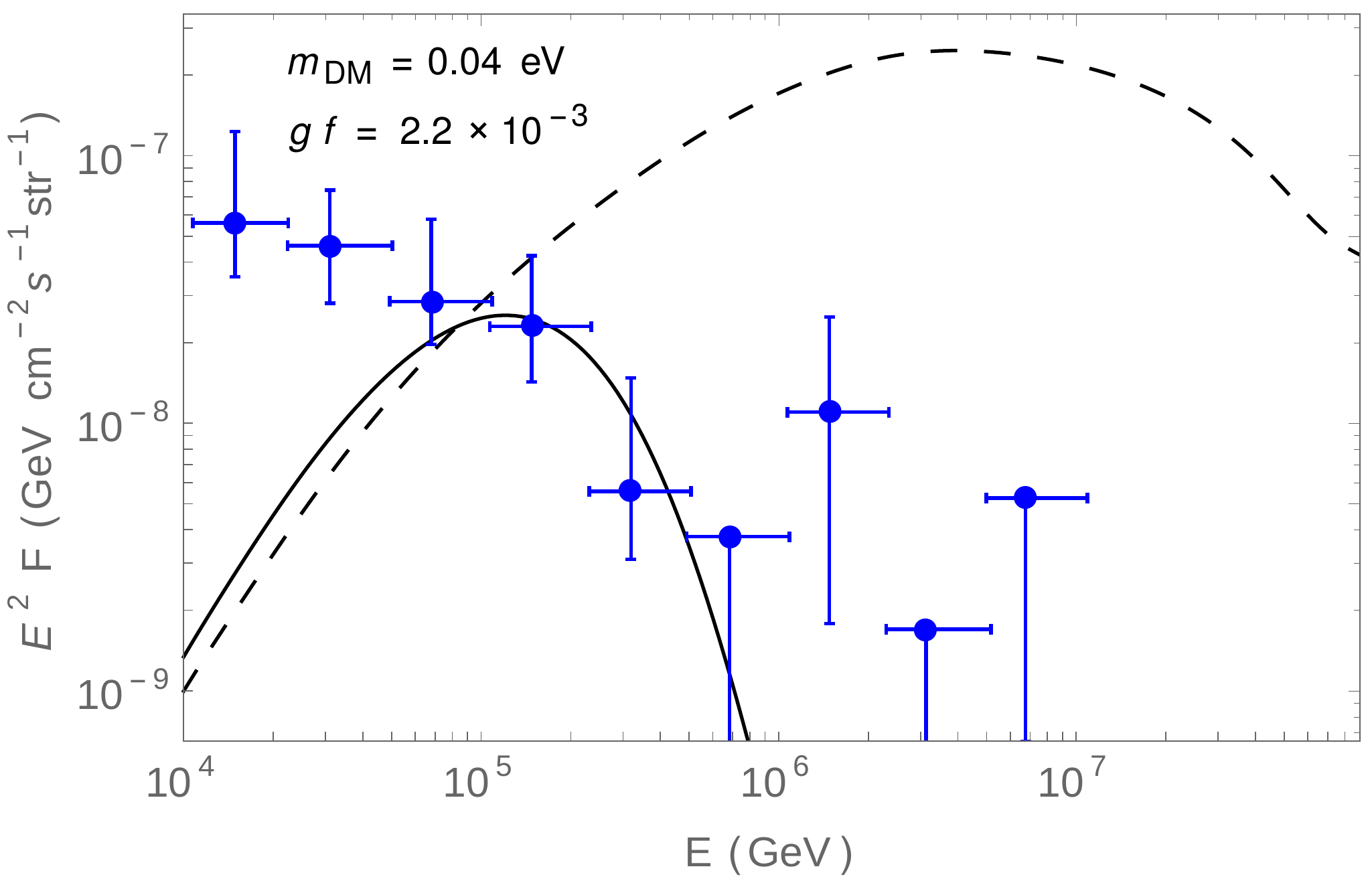}
 \caption{
  Attenuation of the diffuse neutrino flux for the AGN core model S05. The dashed and solid black lines represent the original flux and the flux degraded by $\nu$--DM interactions. Both are multiplied by $E^2$ while plotting.}
  \label{fig:AGN}
   \end{center}
 \end{figure}

So far to demonstrate the potential of the mechanism to explain various features of the observed spectra, we have used fixed power fluxes. We now turn our attention to flux expected in Nature from specific sources, such as AGNs. 
Many of the AGN core and jet models predict an unacceptably large flux at energies more than a PeV, which may be reconciled in the scenario proposed here as the $\nu$--DM cross-section is significant around a PeV or so depending on the parameter space. Neutrinos of energies more than a PeV gets absorbed leaving a low energy excess. In  Fig.~\ref{fig:AGN} we illustrate this with a radio-quiet AGN core model (S05) by Stecker~\cite{Stecker:2013fxa}. The resulting flux peaks at much lower energies, around 100~TeV, and is consistent with observations. Considering AGNs which are farther away, as pointed out earlier, the neutrinos may experience further degradation in energy, which might account for the low energy excess of neutrinos even less than 60\;TeV. Although we have illustrated this with a specific AGN source, the low energy excess might result from regeneration of neutrinos emanating from other sources as well.  

The UHE neutrinos can interact with the relic neutrinos on their way to the Earth mediated by the same $Z^\prime$.  The number density of the relic neutrinos is much less in comparison to the ultralight DM considered above. Hence, the effect of such interactions are negligible away from the $s$-channel resonance. Around the resonance the enhanced cross-section can compensate for the small number density of relic neutrinos. But the width of this resonance, translated into the energy of the incoming neutrino, is around ${\cal O}(100)$\;MeV, implying that the resonance is too sharp to lead to any appreciable neutrino absorption owing to the paucity of UHE neutrinos.

 As shown in this letter, the proposed mechanism can account for an early cut-off immediately after a PeV, explain a dip above 400 TeV and can even generate a low energy excess. For all these different fluxes at different energies are required. This is likely as different dynamics at various sources produce spectra with different power-law behavior with the energy of the neutrino. A simultaneous explanation of various features using the same mechanism may require a multitude of ultralight BEC DM candidates which may easily be envisaged in a multipartite DM model with an elaborate dark sector. It is also quite possible that these DM candidates interact with UHE neutrinos differently, governed by a richer mode of interactions, but following the basic mechanism outlined in this letter. As demonstrated in Fig.~\ref{fig:AGN}, $\nu$--DM interactions can be responsible in suppressing neutrino flux more than a PeV. This helps in reconciling astronomical objects like AGNs as a plausible source of cosmic neutrinos, in particular those models of AGNs which predict large neutrino flux at very high energies.

{\bf\emph{Acknowledgements:}} This work is supported by the Department of Science and Technology, India {\it via} SERB grant EMR/2014/001177 and DST-DAAD grant INT/FRG/DAAD/P-22/2018.


\begin{thebibliography}{99}
\bibitem{IceCube:2018dnn} 
  M.~G.~Aartsen {\it et al.}, 
  Science {\bf 361}, no. 6398, eaat1378 (2018).
  
\bibitem{Murase:2010gj} 
  K.~Murase, and J.~F.~Beacom,
  Phys.\ Rev.\ D {\bf 81}, 123001 (2010).
   
   
\bibitem{Hu:2000ke} 
  W.~Hu, R.~Barkana, and A.~Gruzinov,
  Phys.\ Rev.\ Lett.\  {\bf 85}, 1158 (2000).

\bibitem{Das:2014agf} 
  S.~Das, and R.~K.~Bhaduri,
  Class.\ Quant.\ Grav.\  {\bf 32}, no. 10, 105003 (2015).
  
\bibitem{spsksr2}
S.~Pandey, S.~Karmakar, and S.~Rakshit, to be communicated.

\bibitem{Naumov:1998sf} 
  V.~A.~Naumov, and L.~Perrone,
  Astropart.\ Phys.\  {\bf 10}, 239 (1999).

  
\bibitem{Rakshit:2006yi} 
  S.~Rakshit, and E.~Reya,
  Phys.\ Rev.\ D {\bf 74}, 103006 (2006).
  
\bibitem{Blum:2014ewa} 
  K.~Blum, A.~Hook, and K.~Murase,
  arXiv:1408.3799 [hep-ph].
  
\bibitem{AlbrechtKarle} 
A.~Karle, Talk presented at La Palma~2018 on behalf of IceCube Collaboration. 
  
  
\bibitem{Stecker:2013fxa} 
  F.~W.~Stecker,
  Phys.\ Rev.\ D {\bf 88}, no. 4, 047301 (2013).
   
\end{thebibliography}
\end{document}